\documentclass[5p]{elsarticle}

\usepackage{lineno,hyperref}
\usepackage[T1,T2A]{fontenc}
\usepackage{amsmath}
\usepackage{amssymb}
\usepackage{makecell}
\usepackage{array}
\newcolumntype{?}{!{\vrule width 1pt}}

\usepackage{booktabs}
\usepackage{color}

\hypersetup{colorlinks,allcolors=black}

\begin{document}

\begin{frontmatter}

\title{High-precision $Q$-value measurement and nuclear matrix elements \\for the double-$\beta$ decay of $^{98}$Mo}

\author[a]{D.A.~Nesterenko\corref{mycorrespondingauthor}}
\cortext[mycorrespondingauthor]{Corresponding author}
\ead{dmitrii.nesterenko@jyu.fi}

\author[b]{L.~Jokiniemi}
\author[a,c]{J.~Kotila}
\author[a]{A.~Kankainen}
\author[a]{Z.~Ge}
\author[a]{T.~Eronen}
\author[a]{S.~Rinta-Antila}
\author[a]{J.~Suhonen}

\address[a]{University of Jyv\"askyl\"a, P.O. Box 35, FI-40014 University of Jyv\"askyl\"a, Finland}
\address[b]{Department of Quantum Physics and Astrophysics and Institute of Cosmos Sciences, University of Barcelona, 08028 Barcelona, Spain}
\address[c]{Center for Theoretical Physics, Sloane Physics Laboratory, Yale University, New Haven, Connecticut 06520-8120, USA}

\begin{abstract}
Neutrinoless double-beta ($0\nu\beta\beta$) decay and the standard two-neutrino double-beta ($2\nu\beta\beta$) decay of $^{98}$Mo have been studied. The double-beta decay $Q$-value has been determined as $Q_{\beta\beta}=113.668(68)$~keV using the JYFLTRAP Penning trap mass spectrometer. It is in agreement with the literature value, $Q_{\beta\beta}=109(6)$~keV, but almost 90 times more precise. Based on the measured $Q$-value, precise phase-space factors for $2\nu\beta\beta$ decay and $0\nu\beta\beta$ decay, needed in the half-life predictions, have been calculated. Furthermore, the involved nuclear matrix elements have been computed in the proton-neutron quasiparticle random-phase approximation (pnQRPA) and the microscopic interacting boson model (IBM-2) frameworks. Finally, predictions for the $2\nu\beta\beta$ decay are given, suggesting a much longer half-life than for the currently observed cases.
\end{abstract}

\begin{keyword}
\texttt Double-$\beta$ decay \sep Binding energies and masses \sep Mass spectrometers \sep Penning trap \sep Nuclear matrix elements \sep Phase-space factors
\end{keyword}

\end{frontmatter}


\section{Introduction}

Double-beta ($\beta\beta$) decay is a nuclear process in which two neutrons turn into protons (or vice versa) and two electrons are emitted. In the standard, two-neutrino double-beta ($2\nu\beta\beta$) decay, the emitted electrons are accompanied by two antineutrinos and hence the lepton number is conserved. Such process has already been observed in about a dozen nuclei, where $\beta$ decay is energetically forbidden or very suppressed \cite{Barabash2020}. However, there is also a hypothetical version of $\beta\beta$ decay, namely neutrinoless double-beta ($0\nu\beta\beta$) decay, where only two electrons are emitted. This process violates the lepton-number conservation law of the standard model (SM) of particle physics by two units, since two leptons are created. The process would only be possible if neutrino is a Majorana particle (meaning its own antiparticle) first hypothesized by Ettore Majorana in 1937 \cite{Majorana1937}. The observation of $0\nu\beta\beta$ decay would therefore answer the open questions about beyond-SM physics such as the matter-antimatter symmetry of the Universe \cite{Fukugita1986,Davidson2008} and the nature of neutrinos \cite{Avignone2008,Vergados2012,Engel2017,Dolinski2019,Ejiri2019}.

The neutrinoless mode is under massive searches by several large-scale experiments worldwide \cite{Adams2021,Agostini2020,Anton2019,Alvis2019,Azzolini2019,Arnold2017,Gando2016}, with the most stringent half-life limits given by $t_{1/2}^{0\nu}\gtrsim 10^{26}\,$y, while the measured half-lives of $2\nu\beta\beta$ decay are of the order $t_{1/2}^{2\nu}\sim 10^{18} - 10^{24}\,$y \cite{Barabash2020}. Another intriguing aspect of $0\nu\beta\beta$ decay is that the half-life of the process is inversely proportional to the square of the effective Majorana mass that depends on the neutrino masses. Hence, one could obtain estimates for the neutrino masses (at present, only the differences of the squares are known) from the measured half-lives \cite{Agostini2019,Biller2021}. The next-generation $\beta\beta$-decay experiments are aiming to fully cover the inverted-hierarchy (meaning that the neutrino mass states follow the ordering $m_3<m_1\lesssim m_2$) region of the neutrino masses \cite{Agostini2021}. However, in order to interpret the results, one needs reliable phase-space factors and $\beta\beta$-decay nuclear matrix elements (NMEs), which need to be predicted from nuclear theory. While the phase-space factors can be accurately calculated \cite{Kotila2012}, the present predictions for the $0\nu\beta\beta$-decay NMEs from different theory frameworks disagree by more than a factor of two \cite{Engel2017}.

In the present paper, we study one of the possible $\beta\beta$ emitters, $^{98}$Mo. So far, there has been no direct $Q$-value measurement for the double-beta decay transition between the nuclear ground states $^{98}$Mo$ \rightarrow ^{98}$Ru. The literature $Q$-value, $109(6)$~keV \cite{Wang2021}, has been limited by the uncertainty in the mass value of $^{98}$Ru. It is mainly based on the mass difference between C$_7$H$_{14}$ and $^{98}$Ru measured using a sixteen-inch double-focusing mass spectrometer in 1960s \cite{Damerow1963}. In this work, we determine the $Q$ value by a direct frequency-ratio measurement of singly-charged $^{98}$Mo$^+$ and $^{98}$Ru$^+$ ions in the JYFLTRAP Penning trap \cite{Eronen2012}. In addition, we have measured the $Q$-value for the double-electron capture of $^{96}$Ru and compared it to the high-precision measurement ($\delta m/m\approx 1.4 \times 10^{-9}$) done at the SHIPTRAP Penning trap \cite{Eliseev2011}.

Based on the measured $Q$-value for the $\beta\beta$-decay of $^{98}$Mo, we calculate the phase-space factors for the two-neutrino and the neutrinoless decay modes. Furthermore, we calculate the NMEs for the two decay modes in two different theory frameworks that are well established for calculating the NMEs in medium-heavy to heavy nuclei: proton-neutron quasiparticle random-phase approximation (pnQRPA) \cite{Hyvarinen2015,Simkovic2008} and microscopic interacting boson model (IBM-2) \cite{Barea:2009zza,Barea:2015kwa}. This is the first time the $^{98}$Mo double-beta decay matrix elements are calculated in no-core pnQRPA and IBM-2 frameworks. Due to the low $Q$-value, the $2\nu\beta\beta$-decay has not been measured - hence, we give estimates for both the $2\nu\beta\beta$-decay and $0\nu\beta\beta$-decay half-lives based on the calculated NMEs and phase-space factors.

\section{Double-beta decay}

\subsection{Two-neutrino double-beta decay}

The $2\nu\beta\beta$-decay half-life can be written in the form
\begin{equation}
 [t_{1/2}^{2\nu}]^{-1}=G_{2\nu}\,\left(g_{\rm A}^{\rm eff}\right)^4\,|M^{2\nu}|^2\;,
  \label{eq:2vbb-half-life}   
\end{equation}
where $G_{2\nu}$ is a phase-space factor for the final-state leptons~\cite{Kotila2012} for the two-neutrino mode and $M^{2\nu}$ is the $2\nu\beta\beta$-decay NME. Here $g_{\rm A}^{\rm eff}$ is the effective value of the axial-vector coupling, quenched relative to the free-nucleon value $g_{\rm A}\simeq 1.27$, as found in many different nuclear-structure calculations for the medium-mass and heavy nuclei along the years (see the recent reviews~\cite{Ejiri2019,Suhonen2017}). The NME can be written as
\begin{equation}
M^{2\nu}=M_{\rm GT}^{2\nu}+\Big(\frac{g_{\rm V}}{g_{\rm A}}\Big)^2M_{\rm F}^{2\nu}\;,
\label{eq:m_2v}
\end{equation}
with Gamow-Teller (GT) and Fermi (F) parts and the vector coupling $g_{\rm V}=1.0$ \cite{Hyvarinen2015}.  In the case of $2\nu\beta\beta$ decay, if isospin is a good quantum number, the Fermi matrix elements should identically vanish.
Thus, in both pnQRPA and IBM-2 calculations the Fermi part of the matrix element is forced to be zero in order to restore isospin symmetry, as explained in Section \ref{sec:NMEs}.

\subsection{Neutrinoless double-beta decay}

The $0\nu\beta\beta$-decay half-life can be written as~\cite{Engel2017}
\begin{equation}
[t_{1/2}^{0\nu}]^{-1}=G_{0\nu}\,\left(g_{\rm A}^{\rm eff}\right)^4\,|M^{0\nu}|^2\,\frac{m^2_{\beta\beta}}{m_e^2}\;,
  \label{eq:0vbb-half-life}
\end{equation}
where $G_{0\nu}$ is a phase-space factor for the final-state leptons~\cite{Kotila2012} in the neutrinoless mode,
and $M^{0\nu}$ is the light-neutrino-exchange $0\nu\beta\beta$-decay NME. The effective Majorana mass $m_{\beta\beta}=\sum_i U_{ei}m_i$ (normalized to the electron mass $m_e$) characterizes the lepton-number violation and depends on the neutrino masses $m_i$ and mixing matrix $U$.

The matrix element $M^{0\nu}$ in Eq. \eqref{eq:0vbb-half-life} consists of Gamow-Teller (GT), Fermi (F) and tensor (T) parts and can be written as~\cite{Engel2017}
\begin{equation}
M^{0\nu}=M_{\rm GT}^{0\nu}-\Big(\frac{g_{\rm V}}{g_{\rm A}}\Big)^2M_{\rm F}^{0\nu}+M_{\rm T}^{0\nu}\;.
\label{eq:m_0v}
\end{equation}

\section{Experimental method and results}

The $Q$-value measurements have been performed using the JYFLTRAP Penning trap mass spectrometer \cite{Eronen2012} at the Ion Guide Isotope Separator On-Line (IGISOL) facility \cite{Moore2013}. The ions of interest were separately produced using two electric discharge ion sources, one with natural ruthenium in the IGISOL target chamber \cite{Rahaman2008} and the other with natural molybdenum at the offline ion source station \cite{Vilen2020}. Most of the ions were produced as singly-charged and accelerated to 30 keV. An electrostatic deflector selected ions from one ion source at a time, blocking the ions from the other source. The ions were mass-separated using a 55$^{\circ}$ dipole magnet and the continuous beam with the selected mass number $A$ was injected into a gas-filled radiofrequency quadrupole (RFQ) \cite{Nieminen2001}. The cooled and bunched ion beam after the RFQ was transported to the JYFLTRAP Penning traps placed inside a 7-T superconducting magnet.

In the first (preparation) trap the ions were cooled, centered and additionally purified using a mass-selective buffer gas cooling technique \cite{Savard1991}. In the second (measurement) trap the cyclotron frequency for an ion with mass $m$ and charge $q$ in the magnetic field $B$, given by
\begin{equation} \label{eq:qbm}
\nu_{c} = \frac{1}{2\pi}\frac{q}{m}B,
\end{equation}
was measured employing the phase-imaging ion-cyclotron-resonance (PI-ICR) technique \cite{Eliseev2013, Eliseev2014, Nesterenko2018}. 


The ion's cyclotron frequency $\nu_c$ was determined as a sum of its radial-motion frequencies in the trap, a magnetron frequency $\nu_-$ and a modified cyclotron frequency $\nu_+$. The measurements followed the scheme described in Ref.~\cite{Eliseev2014}. Two excitation patterns were applied alternately in order to determine the accumulated magnetron and cyclotron phases of the ion motion. After injecting the ions into the measurement trap, the coherent component of the magnetron motion was reduced by applying 600-$\mu$s dipolar radiofrequency (rf) pulse at the magnetron frequency $\nu_-$. Then, the cyclotron motion of the ions was excited to an amplitude of about 1 mm through application of a 100-$\mu$s dipolar rf pulse at the modified cyclotron frequency $\nu_+$. After the excitation, the ion's cyclotron motion was converted into the magnetron motion via a 2-ms quadrupolar rf pulse at the frequency close to the cyclotron frequency $\nu_c$. The ions accumulated the magnetron-motion phase during the phase accumulation time $t_{acc}$ of free rotation and were then extracted from the trap. For the measurement of the cyclotron-motion phase the ions accumulated the cyclotron phase after $\nu_+$-pulse for the phase accumulation time $t_{acc}$, which was followed by a conversion pulse applied before the extraction from the trap. The radial motion phase of the ions extracted from the trap was projected onto a position-sensitive detector (microchannel plate detector with a delay line anode).

\begin{table*}
\centering
\caption{\label{tab:results} The weighted means ($\bar{R}$) of the measured frequency ratios $R=\nu_c(\textnormal{daughter})/\nu_c(\textnormal{parent})$ and the corresponding $Q$-values for the studied transitions. The literature $Q$-values and differences to the literature values are also given. }
\begin{tabular}{lllll}
\hline\noalign{\smallskip}
Transition & Frequency ratio $\bar{R}$ & $Q$-value (keV) & Lit. $Q$-value (keV) & Difference (keV) \\
\hline\noalign{\smallskip}
$^{98}$Mo$ \rightarrow ^{98}$Ru & 1.000 001 246 39(74) & 113.668(68) & 109(6)  \:\:\:\:\:\:\:\: \cite{Wang2021} & 4.7(60) \\
$^{96}$Ru$ \rightarrow ^{96}$Mo & 1.000 030 386 86(55) & 2714.583(50) & 2714.51(13)  \cite{Eliseev2011} & 0.073(139)
\\
\noalign{\smallskip}\hline
\end{tabular}
\end{table*} 

\begin{figure}[htb]
\centering
\includegraphics[width=0.5\textwidth]{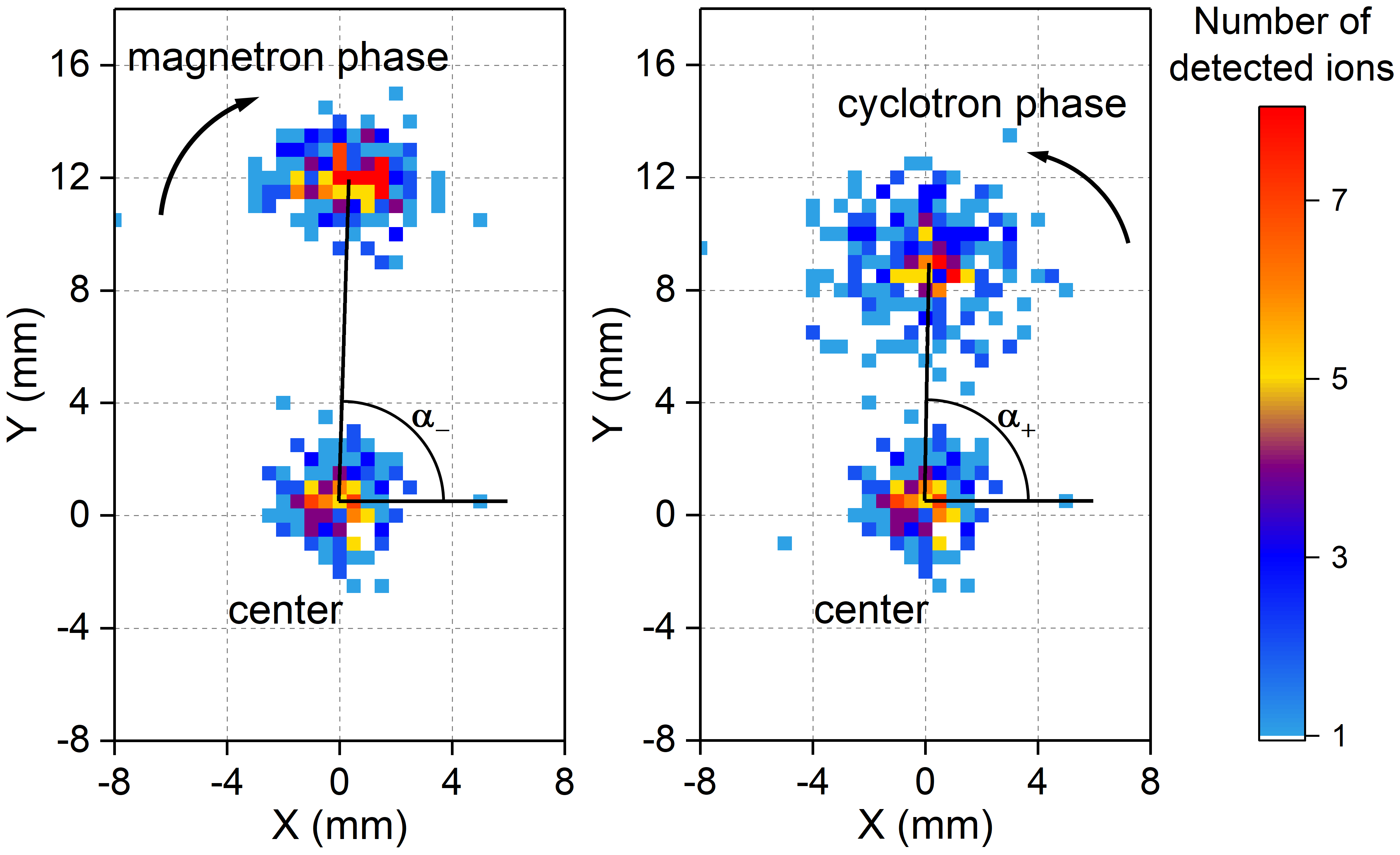}
\caption{Projection of the trap center and accumulated phase spots for $^{98}$Mo$^+$ ions on the position-sensitive MCP detector in a single 4.5-min cyclotron frequency measurement with the PI-ICR method. The phase accumulation time $t_{acc}$ was about of 500~ms.}
\label{fig:PIICR}
\end{figure}

The positions of the magnetron and cyclotron phase images on the detector, defined by the polar angles $\alpha_-$ and $\alpha_+$, correspondingly, with respect to the trap center, were chosen such that the angle $\alpha_c = \alpha_+ - \alpha_-$ did not exceed a few degrees. It is required to minimize the systematic shifts due to the distortion of the projection on the detector and reduce the influence from the conversion of the cyclotron motion to the magnetron motion \cite{Eliseev2014}. The cyclotron frequency is determined from the angle between two phase images as:
\begin{equation} \label{eq:alpha}
\nu_{c}= \nu_{-} + \nu_{+} = \frac{\alpha_c + 2 \pi n} {2 \pi t_{acc}},
\end{equation}
where $n$ is the full number of revolutions, which the studied ions would perform in a magnetic field $B$ in absence of electric field during a phase accumulation time $t_{acc}$.


The phase spots and the center spot were alternately accumulated during a single 4.5-min cyclotron frequency measurement (see Fig.~\ref{fig:PIICR}). About 300 ions were collected for each spot. Any residual magnetron and cyclotron motion could lead to shifts of the phase position. To eliminate these effects, the start time of the cyclotron excitation was repeatedly scanned over a magnetron period ($\approx$600~$\mu$s) and the start time of the extraction pulse was scanned over a cyclotron period ($\approx$0.9~$\mu$s).

The cyclotron frequencies of the parent nuclide $\nu_c^{p}$ and the daughter nuclide $\nu_c^{d}$ were alternately measured changing every 4.5 minutes. The frequency $\nu_c^{d}$ measured before and after the $\nu_c^{p}$ measurement, was linearly interpolated to the time of the $\nu_c^{p}$ measurement and a single cyclotron frequency ratio $R_i = \nu_c^{d} / \nu_c^{p}$ was determined. The systematic uncertainty due to non-linear changes of the magnetic field was negligible compared to the achieved statistical uncertainty \cite{Nesterenko2021}. The final cyclotron frequency ratio $\bar{R}$ was calculated as a weighted mean of $R_i$. The ions of the parent and daughter nuclides were measured in similar conditions to minimize a possible systematic shift of the frequency ratio due to imperfections of the measurement trap. Mass-dependent systematic effects are negligible compared to the statistical uncertainty for mass doublets \cite{Roux2013}. Count-rate class analysis \cite{Kellerbauer2003,Roux2013} was performed and no correlations between the frequency ratios and the number of detected ions per bunch were observed. Up to 5 ions/bunch were taken into account in the analysis.

\begin{figure*}[htb]
\centering
\includegraphics[width=0.95\textwidth]{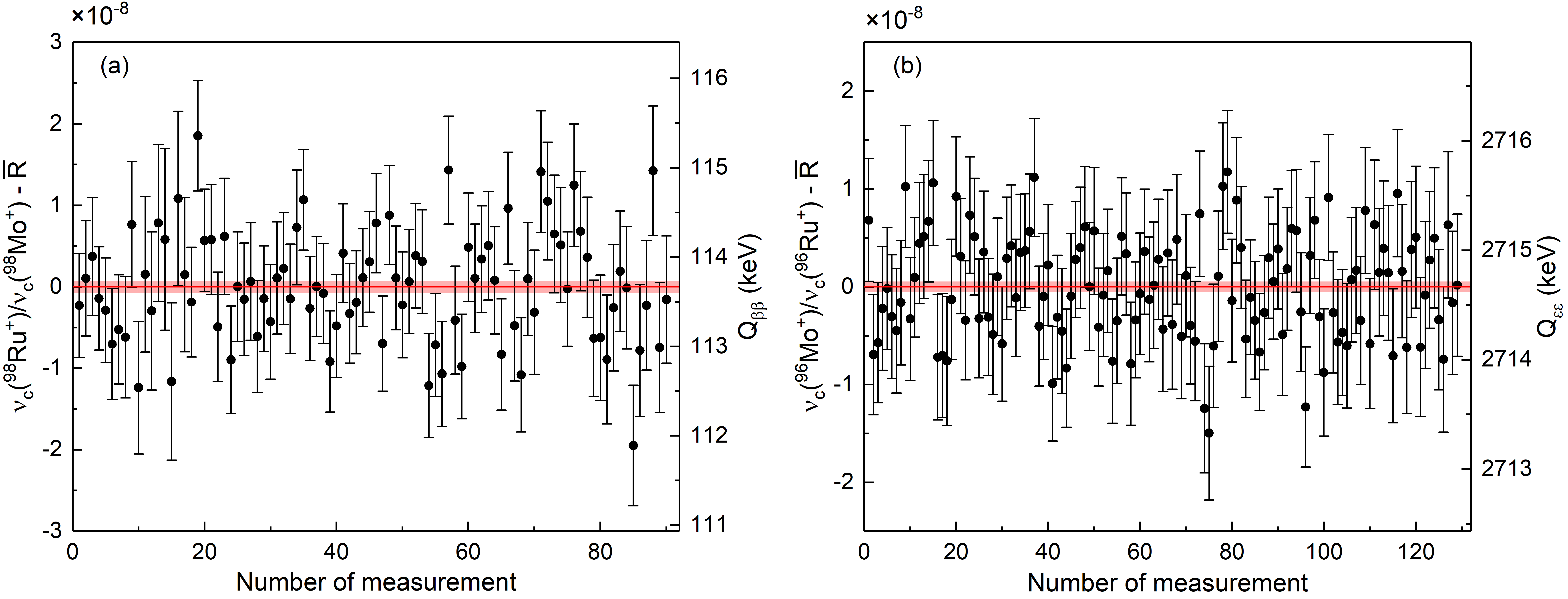}
\caption{Cyclotron frequency ratios (a) $\bar{R} = \nu_c(^{98}$Ru$^+) / \nu_c(^{98}$Mo$^+)$ and (b) $\bar{R} = \nu_c(^{96}$Mo$^+) / \nu_c(^{96}$Ru$^+)$  measured in this work. The red band represents the total 1$\sigma$ uncertainty of the weighted mean frequency ratio $\bar{R}$. For $\bar{R}$ see Table~\ref{tab:results}.}
\label{fig:ratios}
\end{figure*}

The $Q$-value is calculated from the cyclotron frequency ratio as
\begin{equation} \label{eq:Q}
Q = (M_p - M_d)c^2 = (\frac{\nu_c^d}{\nu_c^p} - 1)(M_d - m_e)c^2,
\end{equation}
where $M_p$ and $M_d$ are the atomic masses and $\nu_c^p$ and $\nu_c^d$ the cyclotron frequencies of the parent and daughter nuclides, correspondingly, $m_e$ is the electron mass and $c$ is the speed of light in vacuum. The difference in binding energies of valence electron in Mo and Ru is less than 1 eV \cite{NIST}, and has been neglected. The atomic mass unit used in the analysis is $u$ = 931494.10242(28) keV/$c^2$ \cite{Huang2021}.

The cyclotron frequency ratios $R_i = \nu_c(^{98}\textnormal{Ru}^+) / \nu_c(^{98}\textnormal{Mo}^+)$, measured at JYFLTRAP, are shown in Fig.~\ref{fig:ratios}. The phase accumulation time $t_{acc}$ of about 500~ms was chosen to ensure that the cyclotron spot was not overlapping with any possible isobaric contamination on the detector. The final weighted mean frequency ratio is $\bar{R}=1.00000124639(74)$ resulting in a $Q_{\beta\beta}$-value of 113.668(68) keV. Using the determined $Q_{\beta \beta}$-value of $^{98}$Mo and the mass-excess value of $^{98}$Mo from AME20, -88115.98(17) keV \cite{Wang2021}, we also improve the mass-excess value for $^{98}$Ru considerably, from -88225(6) keV in AME20 \cite{Wang2021} to -88229.65(19) keV.

Similarly, the $Q$-value of double-electron capture in $^{96}$Ru was measured using the PI-ICR technique with the phase accumulation time $t_{acc}$ of about 510~ms. The results are given in Table~\ref{tab:results}. The measured $Q_{\epsilon \epsilon}$-value of $^{96}$Ru, 2714.583(50) keV, is in a good agreement with the SHIPTRAP $Q_{\epsilon \epsilon}$-value, 2714.51(13) keV \cite{Eliseev2011}, and 2.6 times more precise. This measurement provides an additional cross-check of our accuracy with Mo and Ru ions in the studied mass region.


\section{Theory predictions}

\subsection{Phase-space factors}
The key ingredient for the evaluation of phase-space factors (PSF) in single- and double-$\beta$ decay are the electron wave functions. These energy-dependent wave functions are used to form decay-mechanism specific combinations and then integrated over available electron energies up to the end-point energy dictated by the $Q$-value.  A general theory of phase-space factors in $\beta\beta$-decay was developed years ago by Doi \textit{et al.} \cite{10.1143/PTP.66.1739,10.1143/PTP.69.602} following the previous work of Primakoff and Rosen \cite{Primakoff_1959}. It was reformulated by Tomoda \cite{Tomoda_1991} who also presented results for selected nuclei. However, in these earlier calculations approximate expression for the electron wave functions at the nucleus was used. Here we evaluate the PSFs using exact Dirac electron wave functions and including screening by the electron cloud by following the procedure given in Ref. \cite{Kotila2012}. The obtained PSFs for $^{98}$Mo are: $G_{0\nu}=6.18\times 10^{-18}$y$^{-1}$ and $G_{2\nu}=3.71\times 10^{-29}$y$^{-1}$ for the neutrinoless and two-neutrino double-beta decay, respectively.

\subsection{Nuclear matrix elements}
\label{sec:NMEs}
\begin{figure*}
    \centering
    \includegraphics{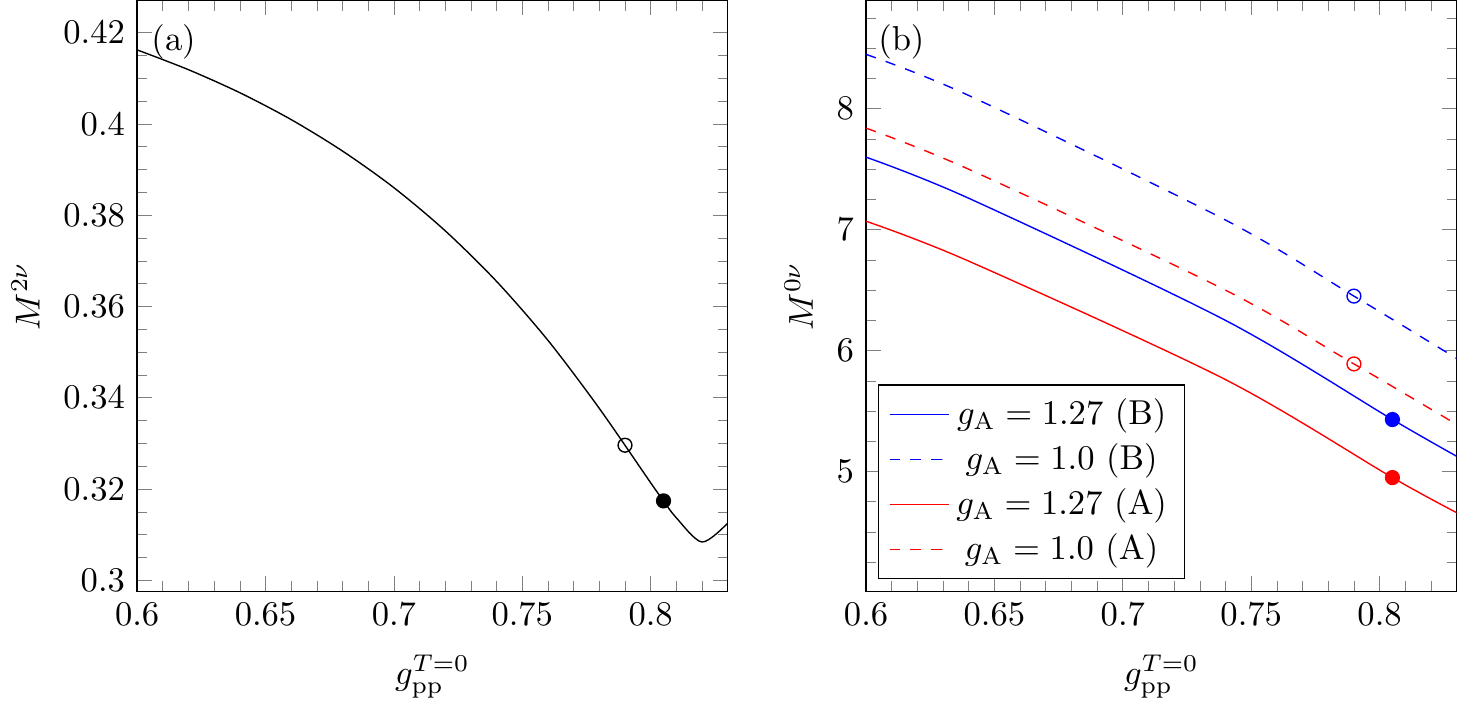}
    \caption{The (a) $2\nu\beta\beta$-decay and (b) $0\nu\beta\beta$-decay NMEs of the transition $^{98}{\rm Mo}(0^+_{\rm g.s})\rightarrow\,^{98}{\rm Ru}(0^+_{\rm g.s})$ as functions of the particle-particle parameter $g_{\rm pp}^{T=0}$ in the pnQRPA framework. The solid (open) circles correspond to the $g_{\rm pp}^{T=0}$ adjusted to the $\log ft$-value of the transition $^{98}{\rm Nb}(1^+_{\rm g.s})\rightarrow\, ^{98}{\rm Mo}(0^+_{\rm g.s})$ with $g_{\rm A}^{\rm eff}=1.27$ ($g_{\rm A}^{\rm eff}=1.0$). 'A' refers to the Argonne and 'B' to the CD-Bonn SRC-parametrization.}
    \label{fig:gpp-2vbb-0vbb}
\end{figure*}

The pnQRPA calculations in the present study are based on the spherical version of pnQRPA with large no-core single-particle bases, similarly as in Refs. \cite{Jokiniemi2018,Jokiniemi2019}. The single-particle bases consist of 25 orbitals - from the lowest $0s_{1/2}$ orbit up to the $0i_{13/2}$ orbit. We take the single-particle energies from a Coulomb-corrected Woods-Saxon potential \cite{Bohr1969}. The quasiparticle spectra, needed in the pnQRPA diagonalization, are obtained by solving the BCS equations using a pairing interaction based on the Bonn-A meson-exchange potential \cite{Holinde1981} for protons and neutrons separately. The interaction is fine-tuned by adjusting the pairing parameters to reproduce the phenomenological pairing gaps. The residual Hamiltonian of the pnQRPA calculation contains two adjustable parameters: the particle-particle $g_{\rm pp}$ and the particle-hole $g_{\rm ph}$ parameters \cite{Suhonen1998}. The particle-hole parameter is adjusted to reproduce the location of the Gamow-Teller giant resonance in $^{98}$Tc. It is a well-known feature \cite{Suhonen1998} that the $\beta$- and $\beta\beta$-decay NMEs are sensitive to the value of particle-particle parameter $g_{\rm pp}$, as demonstrated in the present calculations in Fig. \ref{fig:gpp-2vbb-0vbb}. Here we follow the so-called partial isospin restoration scheme \cite{Simkovic2013}, and divide the parameter into isoscalar ($T=0$) and isovector ($T=1$) parts which multiply the isoscalar and isovector channels of the calculations, respectively. The strength $g_{\rm pp}^{T=1}$ of the isovector channel is then adjusted so that the Fermi part of the $2\nu\beta\beta$-decay NME vanishes. Ideally, the isoscalar strength $g_{\rm pp}^{T=0}$ would then be fixed so that $M_{\rm GT}^{2\nu}$ reproduces the measured $2\nu\beta\beta$-decay half-life, but since it has not been measured for $^{98}$Mo, we adjust $g_{\rm pp}^{T=0}$ to the observed Gamow-Teller transition $^{98}{\rm Nb}(1^+_{\rm g.s})\rightarrow\,^{98}{\rm Mo}(0^+_{\rm g.s})$ with $\log ft=4.72$, instead.

Since the Fermi part of the $2\nu\beta\beta$ NMEs is forced to zero, the $2\nu\beta\beta$ decay runs only through the $1^+$ virtual states of the intermediate nucleus. Hence, the $2\nu\beta\beta$-decay NME is calculated by summing over the $1^+$ states. On the other hand, $0\nu\beta\beta$ decays run through all $J^{\pi}_i$ states in the intermediate nucleus. In the pnQRPA framework, the $0\nu\beta\beta$-decay NMEs are calculated without resorting to the so-called closure approximation by explicitly summing over the intermediate states  (see e.g. \cite{Hyvarinen2015,Simkovic2008}). The wave functions and excitation energies of the states in an odd-odd nucleus are obtained from a pnQRPA diagonalization based on a neighboring even-even reference nucleus \cite{Suhonen2007}. Here, due to the involved two steps of the $\beta\beta$ decay, they are computed in the intermediate odd-odd nucleus $^{98}$Tc by starting from the $^{98}$Mo and $^{98}$Ru reference nuclei. Hence, we obtain two sets of the intermediate states, which is taken into account as an overlap factor in the NMEs \cite{Hyvarinen2015,Suhonen1998}.

Another frequently used model to evaluate $\beta\beta$ NMEs is the microscopic interacting boson model (IBM-2) \cite{ARIMA1977205,iac87}. The method of evaluation is discussed in detail in \cite{Barea:2009zza,Barea:2015kwa}. The logic of the method is to map \cite{OTSUKA19781} the fermion Hamiltonian $H$ onto a boson space and evaluate it with bosonic wave functions.
The single-particle and -hole energies and strengths of interaction were evaluated and discussed in detail in Ref. \cite{PhysRevC.94.034320} where the occupancies
of the single-particle levels were calculated in order to satisfy a twofold goal:
to assess the goodness of the single-particle energies and to check
the reliability of the used wave functions. Both tests are particularly
important in the case of nuclei involved in double beta decay, as
they affect the evaluation of the NMEs and thus their reliability
\cite{eng15}. 

In IBM-2 the isospin is restored by modifying the mapped operator by imposing the condition that $M_{\rm F}^{(2\nu)}=0$. This condition is simply implemented in the calculation by replacing the radial integrals of Appendix A of Ref. \cite{Barea:2009zza} with ones given in Eqs. (9) and (10) in \cite{Barea:2015kwa} that guarantee that the Fermi matrix elements vanish for $2\nu\beta\beta$ decay, as discussed in \cite{Barea:2015kwa}. This replacement also reduces the Fermi matrix elements for $0\nu\beta\beta$ decay by quenching the monopole term from the multipole expansion of the matrix element. Even though the method of isospin restoration is similar in spirit to that of pnQRPA, it is different in practise. 

In the IBM-2 calculations closure approximation is assumed. The main idea behind the closure approximation is to replace the energies of the intermediate states with an average energy, and then the sum over the intermediate states can be removed by using the completeness relation. In \cite{Barea:2013bz} the sensitivity of the $0\nu\beta\beta$ NME to the closure energy ($\sim 10$MeV) is estimated to 
be only 5 \%, owing to the fact that the
momentum of the virtual neutrino is of the order of $100-200$ MeV, i.e., much larger than the typical
nuclear excitations. 

Since the effective value of the axial coupling $g_{\rm A}$ in finite nuclei is under debate \cite{Suhonen2017,Towner1987,Suhonen2019}, we calculate the $2\nu\beta\beta$- and $0\nu\beta\beta$-decay NMEs with two different effective $g_{\rm A}$ values: the free-nucleon value $1.27$ and a standard "shell-model-type" quenched value 1.0. As can be seen from Eq. \eqref{eq:m_2v}, the $2\nu\beta\beta$-decay NME, once the Fermi part is forced to zero, does not depend on $g_{\rm A}$. However, the $g_{\rm A}$-dependence of the NME in the pnQRPA framework stems from the way we adjust the parameter $g_{\rm pp}$, in the present case using the decay rate of a $\beta$-decay transition. The many-body states involved in the $0\nu\beta\beta$-decay NMEs are corrected for the two-nucleon short-range correlations (SRCs) following the so-called CD-Bonn and Argonne parametrizations~\cite{Simkovic2009}. The resulting NMEs for $2\nu\beta\beta$ and $0\nu\beta\beta$ decays are shown in Tables \ref{tab:2vbb-NMEs} and \ref{tab:0vbb-NMEs}, respectively. For $2\nu\beta\beta$ decay we also show the half-lives obtained from Eq. \eqref{eq:2vbb-half-life} with the calculated phase-space factors and NMEs. For comparison, we show the half-lives obtained in projected Hartree-Fock-Bogoliubov (PHFB) \cite{Chandra2005} and in the self-consistent renormalized QRPA (SRQRPA) \cite{Bobyk2000} frameworks with less precise estimates for the phase-space factor. As for $0\nu\beta\beta$-decay, it is hard to give estimates for the half-life, since it depends on the unknown Majorana mass (see Eq. \eqref{eq:0vbb-half-life}). Hence, in Table \ref{tab:0vbb-half-lives} we give estimates for the half-life for a Majorana mass range of $0.01~{\rm eV}<m_{\beta\beta}<0.1~{\rm eV}$, which covers the part of the inverted-hierarchy band of Majorana mass allowed by cosmological searches -- the region the next-generation experiments are interested in.

\begin{table*}
    \caption{The $2\nu\beta\beta$-decay NMEs and the resulting half-lives with different effective $g_{\rm A}$ values for the transition $^{98}{\rm Mo}\rightarrow\,^{98}{\rm Ru}$ calculated in the pnQRPA and IBM-2 frameworks. The half-life results are compared with available other calculations.}
    \centering
    \begin{tabular}{ccccccc}
    \toprule
         $g_{\rm A}^{\rm eff}$ &\multicolumn{2}{c}{$M^{2\nu}$ } &\multicolumn{4}{c}{$t^{2\nu}_{1/2}(10^{29}~{\rm y})$}\\
         \cmidrule(lr){2-3}\cmidrule(lr){4-7}
          &pnQRPA & IBM-2 &pnQRPA &IBM-2 &PHFB \cite{Chandra2005} &SRQRPA \cite{Bobyk2000}\\
         \midrule
         1.27 &0.317 &0.380 &1.031 &0.718 &6.09 &4.06 -- 15.2\\
         1.0  &0.330 &0.380 &2.475 &1.867 &14.87\\
         \bottomrule
    \end{tabular}
    \label{tab:2vbb-NMEs}
\end{table*}

\begin{table*}
    \caption{The $0\nu\beta\beta$-decay NMEs for the transition $^{98}{\rm Mo}\rightarrow\,^{98}{\rm Ru}$ calculated in the pnQRPA and IBM-2 frameworks with different short-range correlations (SRC) and values of $g_{\rm A}^{\rm eff}$. Here $M'^{0\nu}$ refers to the so-called "effective" NME $(g_{\rm A}^{\rm eff}/g_{\rm A})^2M^{0\nu}(g_{\rm A}^{\rm eff})$.}
    \centering
    \begin{tabular}{cccccccccccc}
    \toprule
         & &\multicolumn{5}{c}{pnQRPA}&\multicolumn{5}{c}{IBM-2} \\
         \cmidrule(lr){3-7}\cmidrule(lr){8-12}
         SRC &$g_{\rm A}^{\rm eff}$ &$M_{\rm F}^{0\nu}$ &$M_{\rm GT}^{0\nu}$ &$M_{\rm T}^{0\nu}$ &$M^{0\nu}$ &$M'^{0\nu}$ &$M_{\rm F}^{0\nu}$ &$M_{\rm GT}^{0\nu}$ &$M_{\rm T}^{0\nu}$ &$M^{0\nu}$ &$M'^{0\nu}$\\
         \midrule
         Argonne &1.27 &-1.57 &4.36 &-0.38 &4.95 &4.95 &-0.48 &4.54 &-0.26 &4.58 &4.58\\
         Argonne &1.0 &-1.58  &4.72  &-0.41 &5.89 &3.65 &-0.48 &4.62 &-0.27 &4.82 &2.99\\
         CD-Bonn &1.27 &-1.69 &4.76  &-0.38 &5.43 &5.43 &-0.52 &4.70 &-0.26 &4.76 &4.76\\
         CD-Bonn &1.0 &-1.69  &5.17  &-0.41 &6.45 &4.00 &-0.52 &4.78 &-0.27 &5.04 &3.12\\
         \bottomrule
    \end{tabular}
    \label{tab:0vbb-NMEs}
\end{table*}

\begin{table}
    \caption{The $0\nu\beta\beta$-decay half-lives with different effective $g_{\rm A}$ values for the transition $^{98}{\rm Mo}\rightarrow\,^{98}{\rm Ru}$ calculated in the pnQRPA and IBM-2 frameworks. The ranges correspond to the adopted range of Majorana mass, $0.01~{\rm eV}<m_{\beta\beta}<0.1~{\rm eV}$.}
    \centering
    \begin{tabular}{ccc}
    \toprule
         $g_{\rm A}^{\rm eff}$ &\multicolumn{2}{c}{$t^{0\nu}_{1/2}(10^{29}~{\rm y})$}\\
         \cmidrule(lr){2-3}
          &pnQRPA & IBM-2 \\
         \midrule
         1.27 &0.55 - 66.3 & 0.72 - 77.4\\
         1.0  &1.02 - 122 & 1.66 - 182\\
         \bottomrule
    \end{tabular}
    \label{tab:0vbb-half-lives}
\end{table}

As can be seen from Table \ref{tab:2vbb-NMEs}, the predicted half-lives of the $2\nu\beta\beta$-decay are of the order of $t_{1/2}^{2\nu}\sim 10^{29}\,$y -- much longer than the currently observed half-lives in other nuclei, owing to the low $Q$-value $\sim 100$ keV of the presently discussed transition. For the measured decays the $Q$-values are of the order of $\sim 1$ MeV. The half-lives obtained in the present work are consistently smaller by a factor of $\approx 4-20$ than those obtained in the PHFB and SRQRPA frameworks, mostly due to the larger NMEs obtained in the present work. Contrary to this, the differences between the pnQRPA and IBM-2 predictions are notably smaller, IBM-2 giving some 15 -- 20 \% larger NMEs than pnQRPA.

As for $0\nu\beta\beta$ decay, one can see from Table \ref{tab:0vbb-NMEs} that the pnQRPA-computed effective NMEs $M'^{0\nu}$, obtained with $g_{\rm A}^{\rm eff}=1.27$, are consistently larger by some 10 -- 15\% than the IBM-2-computed ones. With $g_{\rm A}^{\rm eff}=1.0$ the difference is larger, 20 -- 30 \%, owing to the $g_{\rm pp}$-adjustment method of pnQRPA, which partially compensates the quenching effect. The differences between the two calculations stem from the quite different magnitudes of the Fermi NME, which is some 3 times larger in the pnQRPA formalism than in the IBM-2, owing to the quenching of the monopole term from the multipole expansion of the IBM-2 Fermi NME. The pnQRPA- and IBM-2 -computed values of the Gamow-Teller NMEs are, however, quite close to each other. Interestingly, the NMEs obtained within both frameworks are smaller than the NMEs obtained in the PHFB framework \cite{Rath2013}, $M'^{0\nu}(g_{\rm A}^{\rm eff}=1.254)=5.94-7.13$ and $M'^{0\nu}(g_{\rm A}^{\rm eff}=1.0)=4.23-5.13$. 

The $0\nu\beta\beta$-decay half-life predictions in Table \ref{tab:0vbb-half-lives}, obtained from Eq. \eqref{eq:0vbb-half-life} with the NMEs of Table \ref{tab:0vbb-NMEs} and the Majorana mass range $0.01~{\rm eV}<m_{\beta\beta}<0.1~{\rm eV}$, are ranging between $0.55\times 10^{29}$ y and $1.82\times 10^{31}$ y. The smaller IBM-2 NMEs are reflected as slightly longer half-lives, but the ranges obtained in both frameworks are wide due to the uncertainty on the Majorana mass. The lower limits are similar to our half-life predictions for the $2\nu\beta\beta$ decay, but the upper limits are $\sim 100$ times larger. In any case, the half-lives are well beyond the half-life sensitivities of the current experiments $S_{1/2}\approx 10^{25}-10^{26}$ y (for other nuclei).

\section{Discussion}

We have determined the $Q$-value for the double-beta decay of $^{98}$Mo directly for the first time using Penning-trap mass spectrometry. The obtained $Q_{\beta \beta}$ = 113.668(68) keV agrees with the $Q_{\beta \beta}$-value given in AME2020, 109(6) keV \cite{Wang2021}, but is almost 90 times more precise. Based on the measured $Q$-value, the phase-space factors for the two-neutrino and neutrinoless double-beta-decay modes were computed. Furthermore, the nuclear matrix elements, involved in the half-life expressions of these decay modes, were calculated in the pnQRPA and IBM-2 frameworks. Within both frameworks we take the isospin restoration into account by forcing the Fermi matrix element of the $2\nu\beta\beta$ decay to vanish. 

The presently obtained $2\nu\beta\beta$ half-lives are consistently smaller than those previously obtained in the PHFB \cite{Rath2013} or in the SQRPA \cite{Bobyk2000} framework, mostly due to larger NMEs obtained in pnQRPA and IBM-2. On the other hand, the differences between the pnQRPA and IBM-2 values are relatively small, IBM-2 giving some 15 -- 20 \% larger NMEs than pnQRPA. As for $0\nu\beta\beta$ decay, the NMEs obtained in pnQRPA and IBM-2 are consistently smaller than those obtained in the PHFB framework \cite{Rath2013}, but pnQRPA predicts some 10 -- 30 \% larger NMEs, depending on the value of $g_{\rm A}^{\rm eff}$ and the SRC-parametrization. This difference largely pertains to the marked differences in the Fermi NME, the Gamow-Teller NMEs being roughly equal. All in all, the predictions given by the two models are in satisfactory agreement, bearing in mind that the theoretical foundations of the two approaches are quite different: The IBM-2 using the closure approximation and a quite restricted single-particle space with renormalized transition operators, and the pnQRPA including explicitly the intermediate virtual states and using a large no-core single-particle space with bare transition operators. According to both models, the half-life of the $2\nu\beta\beta$-decay of $^{98}$Mo, corresponding to the presently obtained $Q$-value, would be notably larger than the currently known experimental half-lives of some other double-beta nuclei.

\section*{Acknowledgements}
This work has been supported by the Finnish Cultural Foundation (Grant No. 00210067) and the  Academy of Finland (Grant Nos. 314733, 320062, 318043, 295207 and 327629). The funding from the European Union’s Horizon 2020 research and innovation programme under grant agreement No. 771036 (ERC CoG MAIDEN) is gratefully acknowledged.

\bibliography{mybibfile}

\end{document}